\documentclass[11pt]{article}
\usepackage{amsmath,amsfonts,latexsym,graphicx,amssymb}
\usepackage[dvips]{epsfig}
\usepackage{amsfonts}
\usepackage{bm}
\date{}

\newcommand{\ot}{{\,\otimes\,}}
\newcommand{{\Cd}}{{\mathbb{C}^d}}

\def\oper{{\mathchoice{\rm 1\mskip-4mu l}{\rm 1\mskip-4mu l}%
{\rm 1\mskip-4.5mu l}{\rm 1\mskip-5mu l}}}
\def\<{\langle}
\def\>{\rangle}
\newtheorem{theorem}{Theorem}
\newtheorem{corrolary}{Corrolary}

\numberwithin{equation}{section}

\begin{document}
\title{\bf A class of positive atomic maps} \author{Dariusz
Chru\'sci\'nski and Andrzej Kossakowski \\
Institute of Physics, Nicolaus Copernicus University,\\
Grudzi\c{a}dzka 5/7, 87--100 Toru\'n, Poland}

\maketitle

\begin{abstract}

We  construct a new class of positive indecomposable maps in the
algebra of $d \times d$ complex matrices. These maps are
characterized by the `weakest' positivity property and for this
reason they are called atomic. This class provides a new reach
family of atomic entanglement witnesses which define important
tool for investigating quantum entanglement. It turns out that
they are able to detect states with the `weakest' quantum
entanglement.

\end{abstract}

\maketitle

\section{Introduction}

One of the most important problems of quantum information theory
\cite{QIT} is the characterization of mixed states of composed
quantum systems. In particular it is of primary importance to test
whether a given quantum state exhibits quantum correlation, i.e.
whether it is separable or entangled. For low dimensional systems
there exists simple necessary and sufficient condition for
separability. The celebrated Peres-Horodecki criterium
\cite{Peres,PPT} states that a state of a bipartite system living
in $\mathbb{C}^2 \ot \mathbb{C}^2$ or $\mathbb{C}^2 \ot
\mathbb{C}^3$ is separable iff its partial transpose is positive.
Unfortunately, for higher-dimensional systems there is no single
universal separability condition.

The most general approach to separability problem is based on the
following observation \cite{Horodeccy-PM}: a state $\rho$ of a
bipartite system living in $\mathcal{H}_A \ot \mathcal{H}_B$ is
separable iff $\mbox{Tr}(W\rho) \geq 0$ for any Hermitian operator
$W$ satisfying $\mbox{Tr}(W P_A \ot P_B)\geq 0$, where $P_A$ and
$P_B$ are projectors acting on $\mathcal{H}_A$ and
$\mathcal{H}_B$, respectively. Recall, that a Hermitian  operator
$W \in \mathcal{B}(\mathcal{H}_A \ot \mathcal{H}_B)$ is an
entanglement witness \cite{Horodeccy-PM,Terhal1} iff: i) it is not
positively defined, i.e. $W \ngeq 0$, and ii) $\mbox{Tr}(W\sigma)
\geq 0$ for all separable states $\sigma$. A bipartite state
$\rho$ living in $\mathcal{H}_A \ot \mathcal{H}_B$ is entangled
iff  there exists an entanglement witness $W$ detecting $\rho$,
i.e. such that $\mbox{Tr}(W\rho)<0$. Clearly, the construction of
entanglement witnesses is a hard task. It is easy to construct $W$
which is not positive, i.e. has at leat one negative eigenvalue,
but it is very difficult to check that $\mbox{Tr}(W\sigma) \geq 0$
for all separable states $\sigma$.

The separability problem may be equivalently formulated in terms
positive maps \cite{Horodeccy-PM}: a state $\rho$ is separable iff
$(\oper \ot \Lambda)\rho$ is positive for any positive map
$\Lambda$ which sends positive operators on $\mathcal{H}_B$ into
positive operators on $\mathcal{H}_A$. Due to the celebrated
Choi-Jamio{\l}kowski \cite{Jam,Choi1} isomorphism there is a one
to one correspondence between entanglement witnesses and positive
maps which are not completely positive: if $\Lambda$ is such a
map, then $W_\Lambda:=(\oper \ot \Lambda)P^+$ is the corresponding
entanglement witness ($P^+$ stands for the projector onto the
maximally entangled state in $\mathcal{H}_A \ot \mathcal{H}_B$).
 Unfortunately, in spite of
the considerable effort, the structure of positive maps (and hence
also the set of entanglement witnesses) is rather poorly
understood [7--44].

Now, among positive linear maps the crucial role is played by
indecomposable maps. These are maps which may detect entangled PPT
states.  Among indecomposable maps there is a set of maps which
are characterized by the `weakest positivity' property: they are
called {\it atomic maps} and they may be used to detect states
with the `weakest' entanglement. The corresponding entanglement
witnesses we call indecomposable and atomic, respectively.

There are only few examples of indecomposable maps in the
literature (for the list see e.g. the recent paper \cite{OSID-W}).
The set of atomic ones is considerably smaller. Interestingly,
Choi first example \cite{Choi1} of indecomposable positive map
turned out to be an atomic one. Recently, Hall \cite{Hall} and
Breuer \cite{Breuer} considered a new family of indecomposable
maps (they were applied by Breuer \cite{Breuer-bis} in the study
of rotationally invariant bipartite states, see also
\cite{Remigiusz}). In this paper we show that these maps are not
only indecomposable but also atomic. Moreover, we show how to
generalize this family to obtain a large family of new positive
maps. We study which maps within this family are indecomposable
and which are atomic.

The paper is organized as follows: in the next Section we
introduce a natural hierarchy of positive convex cones in the
space of (unnormalized) states of bipartite $d \ot d$ quantum
systems and recall basis notions from the theory of entanglement
witnesses and positive maps.  Section \ref{SEC-BH} discusses
properties of the recently introduced indecomposable maps
\cite{Hall,Breuer} and provides the proof that these maps are
atomic.  Finally, Section \ref{NEW} introduces a new class of
indecomposable maps and studies which maps within this class are
atomic. A brief discussion is included in the last section.

\section{Quantum entanglement vs. positive maps}

Let $M_d$ denote a set of $d \times d$ complex matrices and let
$M_d^+$ be a convex set of semi-positive elements in $M_d$, that
is, $M_d^+$ defines a space of (unnormalized) states of $d$-level
quantum system. For any $\rho \in (M_d \ot M_d)^+$ denote by
$\mathrm{SN}(\rho)$  a Schmidt number of $\rho$ \cite{SN}. This
notion enables one to introduce the following family of positive
cones:
\begin{equation}\label{}
    \mathrm{V}_r = \{\, \rho \in (M_d \ot M_d)^+\ |\
    \mathrm{SN}(\rho) \leq r\, \}  \ .
\end{equation}
One has the following chain of inclusions
\begin{equation}\label{V-k}
\mathrm{V}_1  \subset \ldots \subset  \mathrm{V}_d \equiv (M_d \ot
M_d)^+\ .
\end{equation}
Clearly, $\mathrm{V}_1$ is a cone of separable (unnormalized)
states and $V_d \smallsetminus  V_1$ stands for a set of entangled
states. Note, that a partial transposition $(\oper_d \ot \tau)$
gives rise to another family of cones:
\begin{equation}\label{}
     \mathrm{V}^l = (\oper_d \ot \tau)\mathrm{V}_l \ ,
\end{equation}
such that $ \mathrm{V}^1  \subset \ldots \subset  \mathrm{V}^d$.
 One has
$\mathrm{V}_1 = \mathrm{V}^1$, together with the following
hierarchy of inclusions:
\begin{equation}\label{}
    \mathrm{V}_1 = \mathrm{V}_1 \cap \mathrm{V}^1 \subset \mathrm{V}_2 \cap
    \mathrm{V}^2   \subset \ldots \subset \mathrm{V}_d \cap \mathrm{V}^d \
    .
\end{equation}
Note, that $\mathrm{V}_d \cap \mathrm{V}^d$ is a convex set of PPT
(unnormalized) states. Finally, $\mathrm{V}_r \cap \mathrm{V}^s$
is a convex subset of PPT states $\rho$  such that
$\mathrm{SN}(\rho) \leq r$ and $\mathrm{SN}[(\oper_d \ot
\tau)\rho] \leq s$.

Consider now a set of positive maps $\varphi : M_d \longrightarrow
M_d$, i.e. maps such that $\varphi(M_d^+) \subseteq M_d^+$.
Following St{\o}rmer definition \cite{Stormer1}, a positive map
$\varphi$ is $k$-positive iff
\begin{equation}\label{}
    (\oper \ot \varphi)(\mathrm{V}_k) \subset
    (M_d \ot M_d)^+\ ,
\end{equation}
and it is $k$-copositive iff
\begin{equation}\label{}
    (\oper \ot \varphi)(\mathrm{V}^k) \subset
    (M_d \ot M_d)^+\ .
\end{equation}
Denoting by $\mathrm{P}_k$ ($\mathrm{P}^k$) a convex cone of
$k$-positive ($k$-copositive) maps one has the following chains of
inclusions
\begin{equation}\label{P-k}
\mathrm{P}_d \subset \mathrm{P}_{d-1} \subset \ldots \subset
\mathrm{P}_2 \subset \mathrm{P}_1 \ ,
\end{equation}
and
\begin{equation}\label{P=k}
\mathrm{P}^d \subset \mathrm{P}^{d-1} \subset \ldots \subset
\mathrm{P}^2 \subset \mathrm{P}^1 \ ,
\end{equation}
 where $\mathrm{P}_d$ ($\mathrm{P}^d$) stands for a set of completely positive (copositive) maps.

A positive map $\varphi : M_d \longrightarrow M_d$ is {\it
decomposable} iff $\varphi \in P_d + P^d$, that is, $\varphi$ can
be written as $\varphi = \varphi_1 + \varphi_2$, with $\varphi_1
\in P_d$ and $\varphi_2 \in P^d$. Otherwise $\varphi$ is {\it
indecomposable}. Indecomposable maps can detect entangled state
from $V_d \cap V^d \equiv $ PPT, that is, bound entangled states.
Finally, a positive map is {\it atomic} iff $\varphi \notin
\mathrm{P}_2 + \mathrm{P}^2$. The importance of atomic maps
follows from the fact that they may be used to detect the
`weakest' bound entanglement, that is, atomic maps can detect
states from $V_2 \cap V^2$.

Actually, St{\o}rmer definition \cite{Stormer1} is rather
difficult to apply in practice. Using the Choi-Jamio{\l}kowski
isomorphism \cite{Jam,Choi1} we may assign to any linear map
$\varphi : M_d \rightarrow M_d$ the following  operator
$\widehat{\varphi} \in M_d \ot M_d$:
\begin{equation}\label{}
\widehat{\varphi} = (\oper_d \ot \varphi) P^+ \in M_d \ot M_d\ ,
\end{equation}
where $P^+$ stands for (unnormalized) maximally entangled state in
$\Cd \ot \Cd$. If $e_i$ $(i=1,\ldots,d)$ is an orthonormal basis
in $\Cd$, then
\begin{equation}\label{J}
\widehat{\varphi}      = \sum_{i,j=1}^d e_{ij} \ot
\varphi(e_{ij})\ ,
\end{equation}
where $e_{ij} = |i\>\<j|$ defines a basis in $M_d$. It is clear
that if $\varphi$ is a positive but not completely positive map
then the corresponding operator $\widehat{\varphi}$ is an
entanglement witness. Now, the space of linear maps
$\mathcal{L}(M_d,M_d)$ is endowed with a natural inner product:
\begin{equation}\label{}
    (\varphi,\psi) = \mathrm{Tr} \Big( \sum_{\alpha=1}^{d^2}\,
    \varphi(f_\alpha)^* \psi(f_\alpha) \Big)\ ,
\end{equation}
where $f_\alpha$ is an arbitrary orthonormal basis in $M_d$.
Taking $f_\alpha = e_{ij}$, one finds
\begin{eqnarray}\label{}
    (\varphi,\psi) &=& \mathrm{Tr} \Big( \sum_{i,j=1}^{d}\,
    \varphi(e_{ij})^* \psi(e_{ij}) \Big) \ = \  \mathrm{Tr} \Big( \sum_{i,j=1}^{d}\,
    \varphi(e_{ij})\psi(e_{ji}) \Big)\ .
\end{eqnarray}
The above defined inner product is compatible with the standard
Hilbert-Schmidt product in $M_d \ot M_d$. Indeed,  taking
$\widehat{\varphi}$ and $\widehat{\psi}$ corresponding to
$\varphi$ and $\psi$, one has
\begin{equation}\label{}
    (\widehat{\varphi},\widehat{\psi})_{\rm HS} =  \mathrm{Tr} ({\widehat{\varphi}}^{\,*}\widehat{\psi})
\end{equation}
and using (\ref{J}) one easily finds
\begin{equation}\label{}
(\varphi,\psi) = (\widehat{\varphi},\widehat{\psi})_{\rm HS}\ ,
\end{equation}
that is, formula (\ref{J}) defines an inner product isomorphism.
This way one establishes the duality between maps from
$\mathcal{L}(M_d,M_d)$ and operators from $M_d \ot M_d$
\cite{Eom}: for any $\rho \in M_d \ot M_d$ and $\varphi \in
\mathcal{L}(M_d,M_d)$ one defines
\begin{equation}\label{DUAL}
\< \rho, \varphi\> := (\rho,\widehat{\varphi})_{\rm HS} \ .
\end{equation}
In the space of entanglement witnesses $\mathbf{W}$ one may
introduce the following family of subsets $\mathbf{W}_r \subset
M_d \ot M_d$:
\begin{equation}\label{}
\mathbf{W}_r = \{\, W\in M_d \ot M_d\ |\ \mathrm{Tr}(W\rho) \geq
0\ , \ \rho \in \mathrm{V}_r\, \}\ .
\end{equation}
One has
\begin{equation}\label{}
(M_d \ot M_d)^+ \equiv \mathbf{W}_d  \subset \ldots \subset
\mathbf{W}_1   \ .
\end{equation}
Clearly, $\mathbf{W} = \mathbf{W}_1 \smallsetminus \mathbf{W}_d$.
Moreover, for any $k>l$, entanglement witnesses from $\mathbf{W}_l
\smallsetminus \mathbf{W}_k$  can detect entangled states from
$\mathrm{V}_k \smallsetminus  V_l$, i.e. states $\rho$ with
Schmidt number $l < \mathrm{SN}(\rho) \leq k$. In particular $W
\in \mathbf{W}_k \smallsetminus \mathbf{W}_{k+1}$ can detect state
$\rho$ with $\mathrm{SN}(\rho)=k$.

Consider now the following class
\begin{equation}\label{}
    \mathbf{W}_r^s = \mathbf{W}_r + (\oper \ot \tau)\mathbf{W}_s\
    ,
\end{equation}
that is, $W \in \mathbf{W}_r^s$ iff
\begin{equation}\label{}
    W = P + (\oper \ot \tau)Q\ ,
\end{equation}
with $P \in \mathbf{W}_r$ and $Q \in \mathbf{W}_s$. Note, that
$\mathrm{Tr}(W\rho) \geq 0$ for all $\rho \in \mathrm{V}_r \cap
\mathrm{V}^s$. Hence such $W$ can detect PPT states $\rho$ such
that $\mathrm{SN}(\rho) \geq r$ or $\mathrm{SN}[(\oper_d \ot
\tau)\rho] \geq s$. Entanglement witnesses from $\mathbf{W}_d^d$
are called decomposable \cite{optimal}. They cannot detect PPT
states. One has the following chain of inclusions:
\begin{equation}\label{}
    \mathbf{W}_d^d\, \subset\, \ldots\, \subset\, \mathbf{W}^2_2\, \subset\, \mathbf{W}^1_1\,
    \equiv\,    \mathbf{W}\ .
\end{equation}
The `weakest' entanglement can be detected by elements from
$\mathbf{W}_1^1 \smallsetminus \mathbf{W}_2^2$. We shall call them
{\em atomic entanglement witnesses}. It is clear that $W$ is an
atomic entanglement witness if there is an entangled state $\rho
\in \mathrm{V}_2 \cap \mathrm{V}^2$ such that $\mathrm{Tr}(W \rho)
<0$. The knowledge of atomic witnesses, or equivalently atomic
maps,  is crucial: knowing this set we would be able to
distinguish all entangled states from separable ones.

\section{A class of atomic maps of Breuer and Hall} \label{SEC-BH}

Recently Breuer and Hall \cite{Breuer,Hall} analyzed  the
following class of positive maps $\varphi : M_d \longrightarrow
M_d$
\begin{equation}\label{B-H}
\varphi^d_U(X) =  \mbox{Tr}(X)\, \mathbb{I}_d - X - UX^TU^*  \ ,
\end{equation}
where $U$ is an antisymmetric  unitary  matrix in $\mathbb{C}^d$
which implies that $d$ is necessarily even and $d\geq 4$ (for
$d=2$ the above map is trivial $\varphi^d_U(X)=0$). One may easily
add a normalization factor such that
\begin{equation}\label{BH-norm}
\widetilde{\varphi}^d_U = \frac{1}{d-2}\ \varphi^d_U\ ,
\end{equation}
 is unital, that is,
$\widetilde{\varphi}^d_U(\mathbb{I}_d)=\mathbb{I}_d$. The
characteristic feature of these maps is that for any rank one
projector $P$ its image under $\varphi^d_U$ reads as follows
\begin{equation}\label{}
    \varphi^d_U(P) = \mathbb{I}_d - P - Q \ ,
\end{equation}
where $Q$ is again rank one projector satisfying $PQ=0$. Hence
$\varphi^d_U(P)\geq 0$  which proves positivity of $\varphi^d_U$.
It was shown \cite{Breuer,Hall} that these maps are not only
positive but also indecomposable.

Interestingly, maps considered by Breuer and Hall are closely
related to a positive map introduced long ago by Robertson
\cite{Robertson1}--\cite{Robertson4}. The Robertson map $\varphi_R
: M_4 \longrightarrow M_4$ is defined as follows
\begin{equation}\label{}
    \varphi_R\left( \begin{array}{c|c} X_{11} & X_{12} \\ \hline X_{21} & X_{22} \end{array}
\right) = \frac 12 \left( \begin{array}{c|c} \mathbb{I}_2\,
\mbox{Tr} X_{22} &  X_{12} + R(X_{21}) \\ \hline  X_{21} +
R(X_{12}) & \mathbb{I}_2\, \mbox{Tr} X_{11}
\end{array} \right) \ ,
\end{equation}
where $X_{kl} \in M_2$ and $R : M_2 \longrightarrow M_2$ is
defined by
\begin{equation}\label{}
    R(a) = \mathbb{I}_2\, \mbox{Tr}a - a \ ,
\end{equation}
that is, $R$ is nothing but the reduction map. Introducing an
orthonormal basis $(e_1,\ldots,e_4)$ in $\mathbb{C}^4$ and
defining $e_{ij} = |e_i\>\<e_j|$, one easily finds the following
formulae:
\begin{eqnarray}\label{}
    \varphi_R(e_{11}) &=&  \varphi_4(e_{22}) = \frac 12 ( e_{33} +
    e_{44}) \ , \nonumber \\
 \varphi_R(e_{33}) &=&  \varphi_4(e_{44}) = \frac 12 ( e_{11} +
    e_{22}) \ , \nonumber \\
  \varphi_R(e_{13}) &=& \frac 12 (e_{13} + e_{42})\ , \nonumber \\
\varphi_R(e_{14}) &=& \frac 12 (e_{14} - e_{32})\ ,  \\
\varphi_R(e_{23}) &=& \frac 12 (e_{23} - e_{41})\ , \nonumber \\
\varphi_R(e_{24}) &=& \frac 12 (e_{24} + e_{31})\ , \nonumber \\
 \varphi_R(e_{12}) &=&    \varphi_R(e_{34})\ =\ 0 \ . \nonumber
\end{eqnarray}
Note, that the Robertson map is unital, i.e.
$\varphi_R(\mathbb{I}_4)=\mathbb{I}_4$.
\begin{theorem} \label{BH-R}
The normalized Breuer-Hall map $\widetilde{\varphi}^4_U$ in $d=4$
is unitary equivalent to the Robertson map $\varphi_R$, that is
\begin{equation}\label{}
    \widetilde{\varphi}^4_U(X) =  U_1 \varphi_R(U_2^*XU_2)U_1^*\ ,
\end{equation}
for some unitaries $U_1$ and $U_2$.
\end{theorem}
{\bf Proof}: Let us observe that
\begin{equation}\label{}
    \Gamma \varphi_R(X) \Gamma^* = \widetilde{\varphi}^4_0(X) \ ,
\end{equation}
where $\Gamma$ is the following $4\times 4$ unitary matrix
\begin{equation}\label{}
    \Gamma = \left( \begin{array}{c|c}  \mathbb{I}_2 & 0 \\ \hline
    0 & - \mathbb{I}_2  \end{array} \right) \ ,
\end{equation}
and  $\widetilde{\varphi}^4_0$ is a normalized Breuer-Hall map
(\ref{B-H}) corresponding to $4 \times 4$ antisymmetric unitary
diagonal matrix\footnote{Actually, $U_0$ may be multiplied by a
unitary block-diagonal matrix
\[  U_0 \longrightarrow U_\Lambda = \left( \begin{array}{c|c}  e^{i\lambda_1}\mathbb{I}_2 & 0
\\ \hline
      0 & e^{i\lambda_2}\mathbb{I}_2   \end{array} \right) \cdot U_0 \  , \]
but the arbitrary phases $\lambda_1$ and $\lambda_2$ do not enter
the game.}
\begin{equation}\label{}
    U_0 = i\, \mathbb{I}_2 \ot \sigma_2 \ .
\end{equation}
Now, any antisymmetric unitary matrix $U$ may be represented as
\begin{equation}\label{UV}
    U = VU_0V^T\ ,
\end{equation}
for some orthogonal matrix $V$. It shows that a general
Breuer-Hall map $\varphi^4_U$ is unitary equivalent to
$\varphi^4_0$
\begin{equation}\label{}
    \varphi^4_U(X) = V\varphi^4_0(V^TXV)V^T\ ,
\end{equation}
and hence (after normalization) to the Robertson map
\begin{equation}\label{}
    \widetilde{\varphi}^4_U(X) = (V\Gamma)\varphi_R(V^TXV)(V\Gamma)^T\ ,
\end{equation}
with $U_1 = V\Gamma$ and $U_2 = V$. \hfill $\Box$

Note, that for $V = \mathbb{I}_4$, one obtains
\begin{eqnarray}  \label{}
\widetilde{\varphi}^4_{\mathbb{I}}(e_{ii}) &=&  \varphi_R(e_{ii})   \ ,\\
\widetilde{\varphi}^4_{\mathbb{I}}(e_{ij}) &=& - \varphi_R(e_{ij})
\ , \ \ \ i \neq j \ ,
\end{eqnarray}

It was already shown by Robertson \cite{Robertson3} that
$\varphi_R$ is indecomposable. However, it turns out that one may
prove the following much stronger property:

\begin{theorem}
Robertson map $\varphi_R$ is atomic.
\end{theorem}
{\bf Proof}: to prove atomicity of $\varphi_R$ one has to
construct a PPT state $\rho \in (M_4 \ot M_4)^+$ such that: 1)
both $\rho$ and its partial transpose $\rho^\tau$ are of Schmidt
rank two, and 2) entanglement of $\rho$ is detected by the
corresponding entanglement witness \[ W_R = (\oper \ot
\varphi_R)P^+_4 = \sum_{i,j=1}^4 e_{ij} \ot \varphi_R(e_{ij})\ .
\]
One easily finds
\begin{equation}\label{}
 W_R = \frac 12 \left( \begin{array}{cccc|cccc|cccc|cccc}
 \cdot& \cdot& \cdot& \cdot& \cdot& \cdot& \cdot& \cdot& \cdot& \cdot& 1& \cdot& \cdot& \cdot& \cdot& 1\\
 \cdot& \cdot& \cdot& \cdot& \cdot& \cdot& \cdot& \cdot& \cdot& \cdot& \cdot& \cdot& \cdot& \cdot& \cdot& \cdot\\
 \cdot& \cdot& 1& \cdot& \cdot& \cdot& \cdot& \cdot& \cdot& \cdot& \cdot& \cdot& \cdot& -1& \cdot& \cdot\\
 \cdot& \cdot& \cdot& 1& \cdot& \cdot& \cdot& \cdot& \cdot& 1& \cdot& \cdot& \cdot& \cdot& \cdot& \cdot  \\ \hline
 \cdot& \cdot& \cdot& \cdot& \cdot& \cdot& \cdot& \cdot& \cdot& \cdot& \cdot& \cdot& \cdot& \cdot& \cdot& \cdot \\
 \cdot& \cdot& \cdot& \cdot& \cdot& \cdot& \cdot& \cdot& \cdot& \cdot& 1& \cdot& \cdot& \cdot& \cdot& 1 \\
 \cdot& \cdot& \cdot& \cdot& \cdot& \cdot& 1& \cdot& \cdot& \cdot& \cdot& \cdot& 1& \cdot& \cdot& \cdot \\
 \cdot& \cdot& \cdot& \cdot& \cdot& \cdot& \cdot& 1& -1& \cdot& \cdot& \cdot& \cdot& \cdot& \cdot& \cdot  \\ \hline
 \cdot& \cdot& \cdot& \cdot& \cdot& \cdot& \cdot& -1& 1& \cdot& \cdot& \cdot& \cdot& \cdot& \cdot& \cdot \\
 \cdot& \cdot& \cdot& 1& \cdot& \cdot& \cdot& \cdot& \cdot& 1& \cdot& \cdot& \cdot& \cdot& \cdot& \cdot \\
 1& \cdot& \cdot& \cdot& \cdot& 1& \cdot& \cdot& \cdot& \cdot& \cdot& \cdot& \cdot& \cdot& \cdot& \cdot \\
 \cdot& \cdot& \cdot& \cdot& \cdot& \cdot& \cdot& \cdot& \cdot& \cdot& \cdot& \cdot& \cdot& \cdot& \cdot& \cdot \\ \hline
 \cdot& \cdot& \cdot& \cdot& \cdot& \cdot& 1& \cdot& \cdot& \cdot& \cdot& \cdot& 1& \cdot& \cdot& \cdot \\
 \cdot& \cdot& -1& \cdot& \cdot& \cdot& \cdot& \cdot& \cdot& \cdot& \cdot& \cdot& \cdot& 1& \cdot& \cdot \\
 \cdot& \cdot& \cdot& \cdot& \cdot& \cdot& \cdot& \cdot& \cdot& \cdot& \cdot& \cdot& \cdot& \cdot& \cdot& \cdot \\
 1& \cdot& \cdot& \cdot& \cdot& 1& \cdot& \cdot& \cdot& \cdot& \cdot& \cdot& \cdot& \cdot& \cdot& \cdot \end{array}
 \right)\ ,
\end{equation}
where to maintain more transparent form we replace all zeros by
dots. Note, that $W_R$ has single negative eigenvalue `$-1$',
`$0$' (with multiplicity 10) and `$+1$' (with multiplicity 5).
Consider now the following state constructed by Ha \cite{Ha1}:
\begin{equation}\label{Ha}
 \rho_{\rm Ha}\ =\  \frac 17\ \left( \begin{array}{cccc|cccc|cccc|cccc}
 1    & \cdot& \cdot& \cdot& \cdot& \cdot& \cdot& \cdot& \cdot& \cdot& -1& \cdot& \cdot& \cdot& \cdot& \cdot\\
 \cdot& \cdot& \cdot& \cdot& \cdot& \cdot& \cdot& \cdot& \cdot& \cdot& \cdot& \cdot& \cdot& \cdot& \cdot& \cdot\\
 \cdot& \cdot& 1& \cdot& \cdot& \cdot& \cdot& \cdot& \cdot& \cdot& \cdot& \cdot& \cdot& \cdot& \cdot& \cdot\\
 \cdot& \cdot& \cdot& \cdot & \cdot& \cdot& \cdot& \cdot& \cdot& \cdot& \cdot& \cdot& \cdot& \cdot& \cdot& \cdot  \\ \hline
 \cdot& \cdot& \cdot& \cdot& 1 & \cdot& \cdot& \cdot& \cdot& \cdot& \cdot& \cdot& \cdot& \cdot& \cdot& \cdot \\
 \cdot& \cdot& \cdot& \cdot& \cdot& \cdot& \cdot& \cdot& \cdot& \cdot& \cdot& \cdot& \cdot& \cdot& \cdot& \cdot \\
 \cdot& \cdot& \cdot& \cdot& \cdot& \cdot& \cdot& \cdot& \cdot& \cdot& \cdot& \cdot& \cdot& \cdot& \cdot& \cdot \\
 \cdot& \cdot& \cdot& \cdot& \cdot& \cdot& \cdot& 1& 1 & \cdot& \cdot& \cdot& \cdot& \cdot& \cdot& \cdot  \\ \hline
 \cdot& \cdot& \cdot& \cdot& \cdot& \cdot& \cdot& 1& 1& \cdot& \cdot& \cdot& \cdot& \cdot& \cdot& \cdot \\
 \cdot& \cdot& \cdot& \cdot& \cdot& \cdot& \cdot& \cdot& \cdot& \cdot& \cdot& \cdot& \cdot& \cdot& \cdot& \cdot \\
 -1& \cdot& \cdot& \cdot& \cdot& \cdot& \cdot& \cdot& \cdot& \cdot& 1& \cdot& \cdot& \cdot& \cdot& \cdot \\
 \cdot& \cdot& \cdot& \cdot& \cdot& \cdot& \cdot& \cdot& \cdot& \cdot& \cdot& 1& \cdot& \cdot& \cdot& \cdot \\ \hline
 \cdot& \cdot& \cdot& \cdot& \cdot& \cdot& \cdot& \cdot& \cdot& \cdot& \cdot& \cdot& \cdot& \cdot& \cdot& \cdot \\
 \cdot& \cdot& \cdot& \cdot& \cdot& \cdot& \cdot& \cdot& \cdot& \cdot& \cdot& \cdot& \cdot& \cdot& \cdot& \cdot \\
 \cdot& \cdot& \cdot& \cdot& \cdot& \cdot& \cdot& \cdot& \cdot& \cdot& \cdot& \cdot& \cdot& \cdot& \cdot& \cdot \\
 \cdot& \cdot& \cdot& \cdot& \cdot& \cdot& \cdot& \cdot& \cdot& \cdot& \cdot& \cdot& \cdot& \cdot& \cdot& \cdot \end{array}
 \right)\ .
\end{equation}
It turns out \cite{Ha1} that $\rho_{\rm Ha}$ is PPT, and both
$\rho_{\rm Ha}$ and $(\oper \ot \tau)\rho_{\rm Ha}$ have Schmidt
rank 2.  One easily finds
\begin{equation}\label{Ha-R}
\mbox{Tr}(W_R \rho_{\rm Ha}) = -1/14<0\ ,
\end{equation}
which  proves atomicity of $\varphi_R$.\footnote{Note, that
$\rho_{\rm Ha}$ is trivially extended from the following state in
$3 \ot 3$:
\begin{equation}\label{}
 \frac 17\ \left( \begin{array}{ccc|ccc|ccc}
 1 &  \cdot& \cdot& \cdot& \cdot& \cdot& \cdot& -1& \cdot\\
 \cdot& 1 &\cdot& \cdot& \cdot& \cdot& \cdot& \cdot& \cdot\\
 \cdot& \cdot& \cdot & \cdot& \cdot& \cdot& \cdot& \cdot& \cdot  \\ \hline
 \cdot& \cdot& \cdot& 1 & \cdot& \cdot& \cdot& \cdot& \cdot \\
 \cdot& \cdot& \cdot& \cdot& \cdot& \cdot& \cdot& \cdot& \cdot \\
 \cdot& \cdot& \cdot& \cdot& \cdot& 1& 1 & \cdot& \cdot  \\ \hline
 \cdot& \cdot& \cdot& \cdot& \cdot& 1& 1& \cdot& \cdot \\
 -1& \cdot& \cdot& \cdot& \cdot& \cdot& \cdot& 1& \cdot \\
 \cdot& \cdot& \cdot& \cdot& \cdot& \cdot& \cdot& \cdot& 1
  \end{array} \right)\ ,
\end{equation}
which, therefore, provides an example of a bound entangled state.
} \hfill $\Box$

\begin{corrolary}
The Breuer-Hall map $\varphi^4_U$ is atomic.
\end{corrolary}
{\bf Proof}: using the relation between $\varphi^4_U$ and the
Roberston map $\varphi_R$
\begin{equation}\label{}
    \varphi^4_U(X) = U_1 \varphi_R(U_2^* XU_2)U_1^* \ ,
\end{equation}
let us compute $\mbox{Tr}(\rho W_U)$, where
\begin{equation}\label{}
    W^4_U = (\oper \ot \varphi^4_U)P^+_4\ ,
\end{equation}
and $\rho$ is an arbitrary state in $4 \ot 4$. One obtains
\begin{eqnarray*}\label{}
    \mbox{Tr}(\rho W^4_U) = \frac 14\, \mbox{Tr}\left(\rho\cdot \sum_{i,j=1}^4 e_{ij} \ot
    \varphi^4_U(e_{ij}) \right) = \frac 14\, \mbox{Tr}\left(\rho \cdot\sum_{i,j=1}^4 e_{ij} \ot
    U_1\varphi_R(U_2^* e_{ij} U_2 )U_1^* \right)\ .
\end{eqnarray*}
Now, introducing $\widetilde{e}_i = U_2^* e_i$, one has
\begin{eqnarray}\label{}
    \mbox{Tr}(\rho W_U) &=& \frac 14\, \mbox{Tr}\left(\rho \cdot\sum_{i,j=1}^4 U_2 \widetilde{e}_{ij} U_2^* \ot
    U_1\varphi_R(\widetilde{e}_{ij})U_1^* \right)  \nonumber\\ &
    =& \mbox{Tr} \Big(\rho \cdot (U_2 \ot U_1)(\oper \ot \varphi_R)P^+_4 (U_2 \ot
    U_1)^* \Big) \nonumber\\ &=& \mbox{Tr}\Big( (U_2 \ot U_1)^* \rho (U_2 \ot U_1)
    \cdot W_R\Big) \ .
\end{eqnarray}
Hence, if $\rho_{\rm Ha}$ witnesses atomiticity of $\varphi_R$,
then $(U_2 \ot U_1) \rho_{\rm Ha} (U_2 \ot U_1)^*$ witnesses
atomiticity of $\varphi^4_U$. \hfill $\Box$

The above result may be immediately generalized as follows
\begin{corrolary}
If a positive map $\varphi : \mathcal{B}(\mathcal{H}_1)
\longrightarrow \mathcal{B}(\mathcal{H}_2)$ is atomic, then
$\widetilde{\varphi} : \mathcal{B}(\mathcal{H}_1) \longrightarrow
\mathcal{B}(\mathcal{H}_2)$ defined by
\begin{equation}\label{}
    \widetilde{\varphi}(X) := U_1 \varphi(U_2^* XU_2)U_1^* \ ,
\end{equation}
is atomic for arbitrary unitary operators $U_1$ and $U_2$ ($U_k :
\mathcal{H}_k \longrightarrow \mathcal{H}_k$; $k=1,2$).
\end{corrolary}

\begin{theorem} \label{TH-BHA}
The Breuer-Hall map $\varphi^d_U : M_d \longrightarrow M_d$ with
even $d$  is atomic.
\end{theorem}
{\bf Proof}: let $\Sigma$ be a 4-dimensional subspace in
$\mathbb{C}^d$. It is clear that $U_\Sigma := U|_\Sigma$ gives
rise to the Breuer-Hall map in 4 dimensions
\[  \varphi_{U_\Sigma}^4 : \mathcal{B}(\Sigma) \longrightarrow
\mathcal{B}(U(\Sigma))\ . \]
This map is atomic and hence it is witnessing by a $4 \times 4$
density matrix supported on $\Sigma$, such that $\rho$ is PPT,
Schmidt rank of $\rho$ and its partial transposition equals 2, and
such that $\mbox{Tr}(\rho W_{U_\Sigma}^4) < 0$. Let us extend the
$4 \times 4$ state $\rho$ into the following $d \ot d$ state:
\begin{equation}\label{}
    \widehat{\rho}_{ij,kl} = \left\{ \begin{array}{cc}
\rho_{ij,kl} \ , & \ \ \ i,j,k,l \leq 4 \\
0 & \ \ \ {\rm otherwise} \end{array} \ , \right.
\end{equation}
where we take a basis $(e_1,\ldots,e_d)$ such that $e_1,\ldots,e_4
\in \Sigma$. It is clear that extended $\widehat{\rho}$ is PPT in
$d \ot d$ and Schmidt rank of $\widehat{\rho}$ and $(\oper \ot
\tau)\widehat{\rho}$ equals again 2. Moreover
\begin{equation}\label{}
    \mbox{Tr}( \widehat{\rho} W^d_U) =  \mbox{Tr}( \rho W_{U_\Sigma}^4) < 0 \ ,
\end{equation}
which proves atomicity of $\varphi^d_U$. \hfill $\Box$

Let us observe that $d$ needs not be even. Indeed, let $d \geq  4$
and let $U$ be antisymmetric unitary operator $U : \Sigma
\longrightarrow \Sigma$, where $\Sigma$ denotes an arbitrary
even-dimensional subspace of $\mathbb{C}^d$. One extends $U$ to an
operator $\widehat{U}$ in $\mathbb{C}^d$ by
\begin{equation}\label{}
    \widehat{U}(x,y) = (U x,0) \ ,
\end{equation}
where $x\in \Sigma$ and $y\in \Sigma^\perp$, and hence,
$\widehat{U}$ is still antisymmetric but no longer unitary in
$\mathbb{C}^d$. Finally, let us define
\begin{equation}\label{BH-gen}
    \varphi_{\widehat{U}}^d(X) = \mbox{Tr}(X)\, \mathbb{I}_d - X - \widehat{U}X^T\widehat{U}^*  \ ,
\end{equation}
that is, it acts as the standard Breuer-Hall map on
$\mathcal{B}(\Sigma)$ only. Note, that
\begin{equation}\label{}
 \varphi_{\widehat{U}}^d(\mathbb{I}_d) = (d-2)\mathbb{I}_d +
 P^\perp\ ,
\end{equation}
where $P^\perp$ denotes a projector onto $\Sigma^\perp$.
Therefore, the normalized map reads as follows
\begin{equation}\label{}
 \widetilde{\varphi}_{\widehat{U}}^d(X) = [(d-2)\mathbb{I}_d +
 P^\perp]^{-1/2} \cdot \varphi_{\widehat{U}}^d(X) \cdot [(d-2)\mathbb{I}_d +
 P^\perp]^{-1/2}\ ,
\end{equation}
and has much more complicated form than (\ref{BH-norm}).
\begin{theorem}  \label{BH-arbitrary}
The formula (\ref{BH-gen}) with arbitrary $d\geq 4$  and even
dimensional subspace $\Sigma$ $($with ${\rm dim}\,\Sigma \geq 4)$
defines a positive atomic map.
\end{theorem}
{\bf Proof}:  let $d > {\rm dim}\,\Sigma =2k \geq 4 $. It is clear
that
\begin{equation}\label{}
    \varphi^{2k}_U :=
    \varphi_{\widehat{U}}^d\Big|_{\mathcal{B}(\Sigma)}\ ,
\end{equation}
defines the standard Breuer-Hall map in $\mathcal{B}(\Sigma)$.
Now, due to Theorem \ref{TH-BHA} the map $\varphi^{2k}_U$ is
atomic. If $\rho$ is a $2k \ot 2k$ state living in $\Sigma \ot
\Sigma$ witnessing atomicity of $\varphi^{2k}_U$, then trivially
extended $\widehat{\rho}\,$ in $\,\mathbb{C}^d \ot \mathbb{C}^d$
witnesses atomicity of $\varphi_{\widehat{U}}^d$. \hfill $\Box$

\section{New classes of atomic maps}  \label{NEW}

Now we are ready to propose the generalization of the class of
positive maps considered by Hall \cite{Hall}:
\begin{equation}\label{H1}
\varphi(X) = \sum_{k<l} \sum_{m<n}
\, c_{kl,mn}\, A_{kl} \, X^T\, A_{mn}^*\ ,
\end{equation}
where
\begin{equation}\label{}
    A_{kl} = e_{kl} - e_{lk} \ ,
\end{equation}
with $c_{kl,mn}$ being a $d \times d$ Hermitian matrix. One
example of such a map is a Breuer-Hall one
\begin{equation}\label{}
    \varphi^d_U (X) = \mathbb{I}_d\, \mbox{Tr}\, X - X - UX^TU^*\ ,
\end{equation}
which is shown to be atomic. Moreover, the well know reduction map
\begin{equation}\label{}
    R(X) = \mathbb{I}_d\, \mbox{Tr}\, X - X\  ,
\end{equation}
belongs to (\ref{H1}). This map is completely co-positive and
hence decomposable. Finally, denote by $\varepsilon$ the following
map
\begin{equation}\label{}
    \varepsilon(X) = \mathbb{I}_d\, \mbox{Tr}\, X \ ,
\end{equation}
which is completely positive and does not belong to (\ref{H1}).

Now, let us introduce the new class which is defined by the
following convex combination:
\begin{equation}\label{}
    \phi^U_x(X) = x\varphi^d_U(X) + (1-x)R(X) = \mathbb{I}_d\, \mbox{Tr}\, X - X - x UX^TU^*\
    .
\end{equation}
It is clear that for $x \in [0,1]$ the above formula defines a
 positive map from the class (\ref{H1}). Note, that
if $\mbox{rank}\, U = 2k<d$, then the matrix $[c_{kl,mn}]$
possesses a negative eigenvalue `$1-xk$' for $x$ satisfying
\begin{equation}\label{}
    \frac 1k < x \leq 1\ ,
\end{equation}
and hence $\phi^U_x(X)$ is indecomposable. Similarly, a family
\begin{equation}\label{}
    \psi_y(X) = y\varepsilon(X) + (1-y)R(X) = \mathbb{I}_d\, \mbox{Tr}\, X -
    yX\ ,
\end{equation}
define for $y\in [0,1]$  decomposable maps from (\ref{H1}).
Finally, consider
\begin{equation}\label{chi}
    \chi^U_{x,y}(X) = y\phi^U_x(X) + (1-y)\psi_y(X) = \mathbb{I}_d\, \mbox{Tr}\, X - yX - x UX^TU^*\
    .
\end{equation}
Now, we are going to establish the range of $(x,y) \in [0,1]
\times [0,1]$ for which $\chi^U_{x,y}$ is atomic.

\begin{theorem}  \label{7/4}
A positive map $\chi^U_{x,y}$ is atomic if $ x+y >  7/4$.
\end{theorem}
{\bf Proof:}  let us start with $d=4$ and $\Sigma = \mathbb{C}^4$
and consider $\chi^U_{x,y}$ with $U=\mathbb{I}$:
\begin{equation}\label{chi}
    \chi^\mathbb{I}_{x,y}(X)  = \mathbb{I}_4\, \mbox{Tr}\, X -  yX - x  X^T \    .
\end{equation}
Let $W^\mathbb{I}_{x,y}$ be the corresponding entanglement
witness:
\begin{eqnarray}\label{}
\lefteqn{ W^\mathbb{I}_{x,y} \ = \ (\oper \ot \chi^U_{x,y})P^+_4 \ = \ } && \\
&& \frac 12
   \left( \begin{array}{cccc|cccc|cccc|cccc}
 1-x& \cdot& \cdot& \cdot& \cdot& y-x&  \cdot& \cdot& \cdot& \cdot& -x& \cdot& \cdot& \cdot& \cdot& -x\\
 \cdot& 1-y& \cdot& \cdot& \cdot& \cdot& \cdot& \cdot& \cdot& \cdot& \cdot& \cdot& \cdot& \cdot& \cdot& \cdot\\
 \cdot& \cdot& 1& \cdot& \cdot& \cdot& \cdot& \cdot& \cdot& \cdot& \cdot& \cdot& \cdot& y& \cdot& \cdot\\
 \cdot& \cdot& \cdot& 1& \cdot& \cdot& \cdot& \cdot& \cdot& -y& \cdot& \cdot& \cdot& \cdot& \cdot& \cdot  \\ \hline
 \cdot& \cdot& \cdot& \cdot& 1-y& \cdot& \cdot& \cdot& \cdot& \cdot& \cdot& \cdot& \cdot& \cdot& \cdot& \cdot \\
 y-x& \cdot& \cdot& \cdot& \cdot& 1-x& \cdot& \cdot& \cdot& \cdot& -x& \cdot& \cdot& \cdot& \cdot& -x \\
 \cdot& \cdot& \cdot& \cdot& \cdot& \cdot& 1& \cdot& \cdot& \cdot& \cdot& \cdot& -y& \cdot& \cdot& \cdot \\
 \cdot& \cdot& \cdot& \cdot& \cdot& \cdot& \cdot& 1& y& \cdot& \cdot& \cdot& \cdot& \cdot& \cdot& \cdot  \\ \hline
 \cdot& \cdot& \cdot& \cdot& \cdot& \cdot& \cdot& y& 1& \cdot& \cdot& \cdot& \cdot& \cdot& \cdot& \cdot \\
 \cdot& \cdot& \cdot& -y& \cdot& \cdot& \cdot& \cdot& \cdot& 1& \cdot& \cdot& \cdot& \cdot& \cdot& \cdot \\
 -x& \cdot& \cdot& \cdot& \cdot& -x& \cdot& \cdot& \cdot& \cdot& 1-x& \cdot& \cdot& \cdot& \cdot& y-x \\
 \cdot& \cdot& \cdot& \cdot& \cdot& \cdot& \cdot& \cdot& \cdot& \cdot& \cdot& 1-y& \cdot& \cdot& \cdot& \cdot \\ \hline
 \cdot& \cdot& \cdot& \cdot& \cdot& \cdot& -y& \cdot& \cdot& \cdot& \cdot& \cdot& 1& \cdot& \cdot& \cdot \\
 \cdot& \cdot& y& \cdot& \cdot& \cdot& \cdot& \cdot& \cdot& \cdot& \cdot& \cdot& \cdot& 1& \cdot& \cdot \\
 \cdot& \cdot& \cdot& \cdot& \cdot& \cdot& \cdot& \cdot& \cdot& \cdot& \cdot& \cdot& \cdot& \cdot& 1-y& \cdot \\
 -x& \cdot& \cdot& \cdot& \cdot& -x&  \cdot& \cdot& \cdot& \cdot& y-x& \cdot& \cdot& \cdot& \cdot& 1-x \end{array}
 \right)\ , \nonumber
\end{eqnarray}

 It is easy to show that
\begin{equation}\label{}
    \mbox{Tr}((\Gamma\rho_{\rm Ha}\Gamma^*) W^\mathbb{I}_{x,y}) = \frac{1}{7}\, (7 - 4x-4y)
    \ ,
\end{equation}
where $\rho_{\rm Ha}$ is defined in (\ref{Ha}). Hence, if
$7-4(x+y) < 0$, then $\chi^\mathbb{I}_{x,y}$ is atomic. Now, it is
clear from the proofs of Theorems \ref{BH-R} and
\ref{BH-arbitrary} that the same result applies for arbitrary $d$
and arbitrary $U$. \hfill $\Box$

Similarly, we may find a region in $(x,y)$ square where
$\chi^U_{x,y}$ is indecomposable. One has
\begin{theorem}
A positive map $\chi^U_{x,y}$ is indecomposable if $ x+y >  3/2$.
\end{theorem}
{\bf Proof:} similarly, as in the proof of the previous theorem,
one computes
\begin{equation}\label{3/2}
    \mbox{Tr}((\Gamma\rho_{\rm new}\Gamma^*) W^\mathbb{I}_{x,y}) = \frac{1}{24}\, (24 - 16x-16y)
    \ ,
\end{equation}
where $\rho_{\rm new}$ is defined by
\begin{equation}\label{}
 \rho_{\rm new} = \frac{1}{24} \left( \begin{array}{cccc|cccc|cccc|cccc}
 2& \cdot& \cdot& \cdot& \cdot& \cdot& \cdot& \cdot& \cdot& \cdot& -1& \cdot& \cdot& \cdot& \cdot& -1\\
 \cdot& 2& \cdot& \cdot& \cdot& \cdot& \cdot& \cdot& \cdot& \cdot& \cdot& \cdot& \cdot& \cdot& \cdot& \cdot\\
 \cdot& \cdot& 1& \cdot& \cdot& \cdot& \cdot& \cdot& \cdot& \cdot& \cdot& \cdot& \cdot& 1& \cdot& \cdot\\
 \cdot& \cdot& \cdot& 1& \cdot& \cdot& \cdot& \cdot& \cdot& -1& \cdot& \cdot& \cdot& \cdot& \cdot& \cdot  \\ \hline
 \cdot& \cdot& \cdot& \cdot& 2 & \cdot& \cdot& \cdot& \cdot& \cdot& \cdot& \cdot& \cdot& \cdot& \cdot& \cdot \\
 \cdot& \cdot& \cdot& \cdot& \cdot& 2 & \cdot& \cdot& \cdot& \cdot& -1& \cdot& \cdot& \cdot& \cdot& -1 \\
 \cdot& \cdot& \cdot& \cdot& \cdot& \cdot& 1& \cdot& \cdot& \cdot& \cdot& \cdot& -1& \cdot& \cdot& \cdot \\
 \cdot& \cdot& \cdot& \cdot& \cdot& \cdot& \cdot& 1& 1& \cdot& \cdot& \cdot& \cdot& \cdot& \cdot& \cdot  \\ \hline
 \cdot& \cdot& \cdot& \cdot& \cdot& \cdot& \cdot& 1& 1& \cdot& \cdot& \cdot& \cdot& \cdot& \cdot& \cdot \\
 \cdot& \cdot& \cdot& -1& \cdot& \cdot& \cdot& \cdot& \cdot& 1& \cdot& \cdot& \cdot& \cdot& \cdot& \cdot \\
 -1& \cdot& \cdot& \cdot& \cdot& -1& \cdot& \cdot& \cdot& \cdot& 2 & \cdot& \cdot& \cdot& \cdot& \cdot \\
 \cdot& \cdot& \cdot& \cdot& \cdot& \cdot& \cdot& \cdot& \cdot& \cdot& \cdot& 2 & \cdot& \cdot& \cdot& \cdot \\ \hline
 \cdot& \cdot& \cdot& \cdot& \cdot& \cdot& -1& \cdot& \cdot& \cdot& \cdot& \cdot& 1& \cdot& \cdot& \cdot \\
 \cdot& \cdot& 1& \cdot& \cdot& \cdot& \cdot& \cdot& \cdot& \cdot& \cdot& \cdot& \cdot& 1& \cdot& \cdot \\
 \cdot& \cdot& \cdot& \cdot& \cdot& \cdot& \cdot& \cdot& \cdot& \cdot& \cdot& \cdot& \cdot& \cdot& 2& \cdot \\
 -1& \cdot& \cdot& \cdot& \cdot& -1& \cdot& \cdot& \cdot& \cdot& \cdot& \cdot& \cdot& \cdot& \cdot& 2 \end{array}
 \right)\ ,
\end{equation}
and turns out to be PPT.\footnote{Actually, we originally
constructed $\rho_{\rm new}$ to `beat' (\ref{Ha-R}). One finds
\begin{equation}\label{}
\mbox{Tr}(W_R \rho_{\rm new}) = -1/6\ ,
\end{equation}
which is `much better' that $-1/14$. We conjecture, that
$\rho_{\rm new}$ is `optimal' in the following sense:
\begin{equation}\label{}
    \min_{\rho\in {\rm PPT}} \mbox{Tr}(W_R \rho) = -1/6\ ,
\end{equation}
that is, $\rho_{\rm new}$ minimizes $\mbox{Tr}(W_R \rho)$ among
all PPT states. } It is therefore clear that for $x + y
> 3/2$, that map $\chi^\mathbb{I}_{x,y}$ is indecomposable. Using
the same techniques as in the proof of Theorem \ref{7/4} we prove
that $x + y > 3/2$ guaranties indecomposability for arbitrary $d$
and $U$. \hfill $\Box$

The regions of indecomposability $(x + y > 3/2)$ and of atomicity
$(x+y> 7/4$) are displayed on Figure 1. We stress that these
regions are derived by using  specific  states: $\rho_{\rm new}$
and $\rho_{\rm Ha}$, respectively. It is interesting to look for
other states which are `more optimal' and enable us to enlarge
these regions.

\begin{figure}[t]
\begin{center}
\epsfig{figure=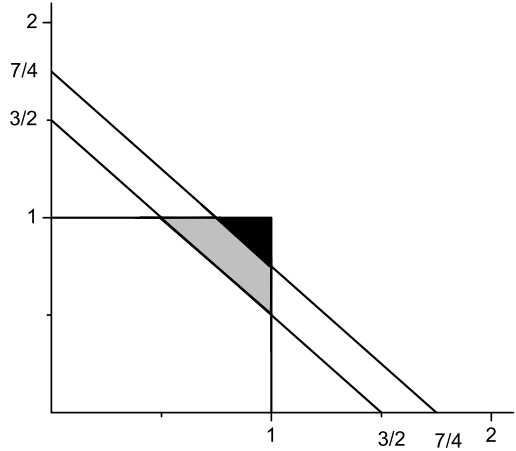,width=0.49\textwidth}
\end{center}
\caption{Regions of indecomposability (gray and black) and of
atomicity (black).}
\end{figure}

 Note, that the same analysis applies for the maps defined by
\begin{equation}\label{H0}
\varphi(X) = \sum_{k<l} \sum_{m<n} \, c_{kl,mn}\, A_{kl} \, X\,
A_{mn}^*\ .
\end{equation}

\section{Conclusions}

We provided a new large class of positive atomic maps in the
matrix algebra $M_d$. These maps generalize a class of maps
discussed recently by Breuer \cite{Breuer} and Hall \cite{Hall}.
The importance of atomic maps follows from the fact that they may
be used to detect the `weakest' bound entanglement, that is,
atomic maps can detect entangled states from $V_2 \cap V^2$. By
duality, these maps provide new class of atomic entangled
witnesses. Note, that if $\varphi$ is atomic and $(\oper \ot
\varphi)\rho \ngeq 0$, then $\rho \in V_2 \cap V^2$ and hence
$\rho$ may be used as a test for atomicity of positive
indecomposable maps. Since we know only few examples of quantum
states belonging to $V_2 \cap V^2$ any new example of this kind is
welcome. It is hoped that new maps provided in this paper find
applications in the study of `weakly' entangled PPT states. For
example in recent papers \cite{PPT-nasza} and \cite{CIRCULANT} we
constructed very general classes of PPT states in $d \ot d$. It
would be interesting to search for entangled states within these
classes by applying our new family of indecomposable and atomic
maps.

\section*{Acknowledgement} This work was partially supported by the
Polish Ministry of Science and Higher Education Grant No
3004/B/H03/2007/33.

\end{document}